\documentclass[onecolumn,fleqn,runningheads]{svjour2}
\usepackage{graphicx}
\usepackage{color}
\usepackage{amsmath}
\usepackage{array}

\usepackage{mathptmx}      
\usepackage{latexsym}
\journalname{Acta Mechanica Sinica}
\usepackage{bm}
\newcommand{\dif}{\mathrm{d}}

\DeclareMathAlphabet{\mathsfsl}{OT1}{cmss}{m}{sl}
\newcommand{\tensor}[1]{\mathsfsl{#1}}

\begin{document}

 \title{An Examination of the Time-Centered Difference Scheme for Dissipative Mechanical Systems from a Hamiltonian Perspective}
\titlerunning{Difference Scheme for Dissipative Systems and Hamiltonian Perspective}

\author{Tianshu Luo \and Yimu Guo}

\institute{Tianshu Luo, Yimu Guo \at
              Institute of Solid Mechanics, Department of Applied Mechanics, Zhejiang University,\\
Hangzhou, Zhejiang, 310027,  P.R.China          
            \email{ltsmechanic@zju.edu.cn}
           \and
           Yimu Guo \at
              Institute of Solid Mechanics, Department of Applied Mechanics, Zhejiang University,\\
Hangzhou, Zhejiang, 310027,  P.R.China          
            \email{guoyimu@zju.edu.cn}
}

\date{Received: date / Accepted: date}
\maketitle
\begin{abstract}
In this paper, we have proposed an approach to observe the time-centered difference scheme for dissipative mechanical systems from a Hamiltonian perspective and
 to introduce the idea of symplectic algorithm to dissipative systems. The dissipative mechanical systems discussed in this paper are finite dimensional. 
This approach is based upon a proposition:
\textcolor{red}{
for any nonconservative classical mechanical system and any initial condition, there exists a conservative one; the two systems share
one and only one common phase curve;  the Hamiltonian of the conservative system is the sum of the total energy of the nonconservative system on the aforementioned phase curve
and a constant depending on the initial condition.}
 Hence, this approach entails substituting an infinite number of conservative systems for a dissipative mechanical system corresponding to varied 
initial conditions. One key way we use to demonstrate these viewpoints is that by the Newton-Laplace principle the nonconservative force can be reasonably 
assumed to be equal to a function of a component of generalized coordinates $q_i$ along a phase curve, such that a nonconservative system can be reformulated 
as countless conservative systems. The advantage of this approach is such that there is no 
need to change the definition of canonical momentum and the motion is identical to that of the original system. Therefore, first we utilize the time-centered 
difference scheme directly to solve the original system, after which we substitute the numerical solution for the analytical solution to construct a 
conservative force equal to the dissipative force on the phase curve, such that we would obtain a substituting conservative system numerically. 
Finally, we use the time-centered scheme to integrate the substituting system numerically. We will find an interesting fact that the latter solution 
resulting from the substituting system is equivalent to that of the former. Indeed, there are two transition matrices within 
time grid points: the first one is unsymplectic and the second symplectic. In fact, the time-centered scheme for dissipative systems
 can be thought of as an algorithm that preserves the symplectic structure of the substituting conservative systems. In addition, via numerical examples
we find that the time-centered scheme preserves the total energy of dissipative systems. According to such behaviors, we might 
explain why some algorithms, e.g., the time-centered Euler scheme, are better than other unsymplectic  algorithms for dissipative systems, 
such that we might choose better algorithms and introduce the idea of this paper to more symplectic algorithms. 
\end{abstract}


\section{Introduction\label{Introduction}}

Feng\cite{Feng1985,Feng1989,Feng1990,Feng1991}, Zhong\cite{Zhong2005,Zhong2009,Gao2009} and Marsden\cite{Marsden1998} have developed a series of symplectic algorithms for conservative systems and
proposed common theories for the construction of symplectic algorithms. 

Feng\cite{Feng1985} has investigated some existing old numerical algorithms from a Hamiltonian 
perspective, such as the simplest symplectic algorithm, the Euler time-centered difference scheme. 
He have explained why for a conservative system the Euler time-centered difference scheme is more accurate than other unsymplectic schemes 
from a Hamiltonian perspective. Because the time-centered difference scheme is derived from Hamilton's equation, the algorithm can preserve symplectic structure and mechanical energy.
\textcolor{red}{The mechanical energy-preserving behavior was proven by Xing\cite{xingyufeng2007}, who found that for dissipative problems this 
algorithm has good mechanical energy-preserving characteristics.} We will state the reason using the Hamiltonian description of dissipative mechanical systems, 
which are finite dimensional in this paper, and we will then apply the idea of symplectic algorithms to the analyses of the time-centered difference scheme for dissipative 
systems. Nevertheless, since Hamilton originated Hamilton equations of motion and Hamiltonian formalism, 
it has been stated in most classical textbooks that the Hamiltonian formalism focuses on solving conservative problems. 

If one needs to apply symplectic algorithms to dissipative systems, one must convert a dissipative system into a Hamiltonian system or find some relationship
between the dissipative system and a conservative one. An attempt\cite{Kane2000} has been made to apply symplectic algorithms directly to dissipative systems. The transition matrix
between phase variable vectors must be unsymplectic. Symplectic algorithms called variational integrators\cite{Kane2000} were derived from discretization of the variational principle for conservative systems
\[
 \delta \int_a^b L(q(t),\dot{q}(t)) \dif t=0,
\]
which is equivalent to Hamilton's equation. Therefore, the algorithms called variational integrators are natively symplectic. But the direct variational integrators 
for dissipative systems were derived from discretization of the Lagrange-d'Alembert principle
\[
 \delta \int_a^b L(q(t),\dot{q}(t)) \dif t+\int_a^b F(q(t),\dot{q}(t)) \delta q \dif t=0,
\]
where $F(q(t),\dot{q}(t))$ is a nonconservative force. Obviously Hamilton's equation cannot be derived from the Lagrange-d'Alembert principle. Zhong\cite{Zhong2009}
 considered that discretization of Hamilton's equations or the Hamiltonian least action variational principle can lead to symplectic transition matrices. 
Therefore, the variational integrators applied directly to dissipative systems cannot be considered as symplectic schemes.

Zhong\cite{Zhong2005} attempted to convert a damping Duffing equation into a conservative system by redefining the position variable $q$ and the canonical momentum $p$
and then applying the time-fem method to this conservative system. His method is to multiply $q$ by the reciprocal of the amplitude decay coefficient such that the
momentum is redefined. The characteristic of the original dissipative system has changed entirely. 

Although several other approaches have been proposed to represent dissipative systems as Hamiltonian formalism, these approaches might not be accepted by 
researchers in geometrical mechanics. For instance, Morrison\cite{Morrison2006593,RevModPhys.70.467} and Salmon\cite{RickSalmon1988} 
focused on the conservative system or some special dissipative systems, e.g. an oscillator with gyroscopic damping. 
Morrison\cite{Morrison2006593} wrote, 'The ideal fluid description is one in which viscosity or other phenomenological terms are neglected. 
Thus, as is the case for systems governed by Newton's second law without dissipation, such fluid descriptions posses Lagrangian and Hamiltonian descriptions.' 
If there had existed an approach appropriate for representing an oscillator with damping as Hamiltonian formalism, 
these researchers would have attempted to extend the Hamiltonian description to other dissipative problems.

Marsden \cite{Marsden2007} and other researchers applied the equations as below to the problem of stability of dissipative systems
\begin{eqnarray}
\dot{p}_i&=&- \frac{\partial H}{\partial q_i}+\bm{F}\left( \frac{\partial{r}}{\partial q_i} \right) \nonumber \\
\dot{q}_i&=& \frac{\partial H}{\partial p_i},
\label{eq:eq2}
\end{eqnarray}
where the position vector $r$ depends on the canonical variable $\lbrace q,p \rbrace$, i.e. $r(q,p)$, $H$ denotes Hamiltonian, 
and $\bm{F} (\partial{r}/ \partial{q_i})$ denotes a generalized force in the direction $i$, $i=1,\dots,n$. 
Marsden considered that Eqs.(\ref{eq:eq2}) was composed of a conservative part and a non-conservative part. Eq.(\ref{eq:eq2}) apparently is 
not a Hamilton's equation but only a representation of dissipative mechanical systems in the phase space.
Although one can utilize the approaches discussed in some papers to convert Eq.({\ref{eq:eq2}}) into a Hamiltonian system, one must first change 
the definition of the canonical momentum of the system. If one uses symplectic algorithms to solve the Hamiltonian system, the numerical
 solution will lose the physical characteristics of the original system, because the phase flow of the original system is different
 from that of the new system. We need a Hamiltonian system that shares common phase flow or solution with the original system. 
But this demand cannot be satisfied, because it conflicts with Louisville's theorem. Therefore, we would have to attempt to find some relationship between 
dissipative systems and conservative ones, such that we can introduce the concept of symplectic algorithms to dissipative problems and 
explain the time-centered difference scheme from Hamiltonian viewpoints.

Based on Eq.(\ref{eq:eq2}), in this paper we will attempt to demonstrate that a dissipative mechanical system shares a single common phase curve with a conservative system. 
In the light of this property, we will propose an approach to substitute a group of conservative systems for a dissipative mechanical system. 
In the following section, we will illustrate the relationship between a dissipative mechanical system and a conservative one.

\section{Relationship between a Dissipative Mechanical System and a Conservative One}

\subsection{A Proposition\label{ObHamiltonian}}
Under general circumstances, the force $\bm{F}$ is a damping force that depends on the variable set $q_1,\cdots,q_n,\dot{q}_1,\cdots,\dot{q}_n$.
$F_i$ denotes the components of the generalized force $\bm{F}$.
\begin{equation}
F_i(q_1,\cdots,q_n,\dot{q}_1,\cdots,\dot{q}_n)=\bm{F}\left( \frac{\partial{r}}{\partial q_i}\right).
\label{eq:inth-1}
\end{equation}
Thus we can reformulate Eq.(\ref{eq:eq2}) as follows:
\begin{eqnarray}
\dot{p}_i&=& - \frac{\partial H}{\partial q_i}
+F_i(q_1,\cdots,q_n,\dot{q}_1,\cdots,\dot{q}_n) \nonumber \\
 \dot{q}_i&=& \frac{\partial H}{\partial p_i}.
\label{eq:inth-2}
\end{eqnarray}
Suppose the Hamiltonian quantity of a conservative system without damping is $\hat{H}$. Thus we
 may write a Hamilton's equation of the conservative system :
\begin{eqnarray}
\dot{p}_i &=& -\frac{\partial {\hat{H}}}{\partial q_i} \nonumber \\
\dot{q}_i &=&\frac{\partial \hat{H}}{\partial p_i}.
\label{eq:inth-3}
\end{eqnarray}
We do not intend to change the definition of momentum in classical mechanics, but we do require that a special solution  
of Eq.(\ref{eq:inth-3}) is the same as that of Eq.(\ref{eq:inth-2}).
We may therefore assume a phase curve $\gamma$ of Eq.(\ref{eq:inth-2}) coincides with 
that of Eq.(\ref{eq:inth-3}). The phase curve $\gamma$ corresponds to an initial condition $q_{i0},p_{i0}$. 
Consequently by comparing Eq.(\ref{eq:inth-2}) and Eq.(\ref{eq:inth-3}), we have
\begin{eqnarray}
\left.\frac{\partial{\hat{H}}}{\partial{q_i}}\right|_{\gamma} &=&
\left.\frac{\partial H}{\partial q_i}\right|_\gamma-\left. F_i(q_1,\cdots,q_n,\dot{q}_1,\cdots,\dot{q}_n)\right|_\gamma \nonumber \\
\left.\frac{\partial{\hat{H}}}{\partial{p_i}}\right|_\gamma&=&
\left.\frac{\partial H}{\partial p_i}\right|_\gamma,
\label{eq:inth-4}
\end{eqnarray}
where $\left.\frac{\partial{\hat{H}}}{\partial{q_i}}\right|_{\gamma},\left.\frac{\partial H}{\partial q_i}\right|_\gamma
,\left.\frac{\partial{\hat{H}}}{\partial{p_i}}\right|_\gamma and \left.\frac{\partial H}{\partial p_i}\right|_\gamma$ denote the values
of these partial derivatives on the phase curve $\gamma$ and $\left.F_i(q_1,\cdots,q_n,\dot{q}_1,\cdots,\dot{q}_n)\right|_\gamma$ denotes 
the value of the force $F_i$ on the phase curve $\gamma$. In classical mechanics the Hamiltonian $H$ of a conservative mechanical 
system is mechanical energy and can be written as:
\begin{equation}
 H=\int_{\gamma}\left(\frac{\partial{H}}{\partial{q_i}}\right)\dif q_i
+\int_{\gamma} \left(\frac{\partial H}{\partial p_i}\right)\dif p_i+const_1,
\label{eq:inth-5}
\end{equation}
where $const_1$ is a constant that depends on the initial condition described above. If $q_i=0,p_i=0$, then $const_1=0$.
The Einstein summation convention has been used this section. Thus an attempt has been made to find $\left.\hat{H} \right|_\gamma$ 
through line integral along the phase curve $\gamma$ of the dissipative system
\begin{eqnarray}
\int_{\gamma}\frac{\partial{\hat{H}}}{\partial{q_i}}\dif q_i
&=&\int_{\gamma}\left[\frac{\partial H}{\partial q_i}
-F_i(q_1,\cdots,q_n,\dot{q}_1,\cdots,\dot{q}_n)\right]\ \dif q_i \nonumber \\
 \int_{\gamma} \frac{\partial \hat{H}}{\partial p_i}\dif p_i
&=&\int_{\gamma} \frac{\partial H}{\partial p_i}\dif p_i.
\label{eq:inth-6}
\end{eqnarray}
Analogous to Eq.(\ref{eq:inth-5}), we have
\begin{equation}
 \left.\hat{H}\right|_{\gamma}=\int_{\gamma}\frac{\partial{\hat{H}}}{\partial{q_i}}\dif q_i
+\int_{\gamma} \frac{\partial{\hat{H}}}{\partial p_i}\dif p_i+const_2,
\label{eq:inth-7}
\end{equation}
where $const_2$ is a constant which depends on the initial condition.
Substituting Eq.(\ref{eq:inth-5})(\ref{eq:inth-6}) into Eq.(\ref{eq:inth-7}), we have
\begin{equation}
 \left.\hat{H}\right|_\gamma=H-\int_{\gamma}F_i(q_1,\cdots,q_n,\dot{q}_1,\cdots,\dot{q}_n)\dif q_i+const.
\label{eq:inth-8}
\end{equation}
where $const=const_2-const_1$, and $H=\left.H\right|_{\gamma}$ because $H$ is mechanical energy of the nonconservative system(\ref{eq:inth-2}). 
According to the physical meaning of Hamiltonian, $const_1$, $const_2$ and $const$ are added into Eq.(\ref{eq:inth-5})(\ref{eq:inth-7})(\ref{eq:inth-8})
respectively such that the integral constant vanishes in the Hamiltonian quantity. 
Arnold\cite{Arnold1997} had presented the Newton-Laplace principle of determinacy as, 
'This principle asserts that the state of a mechanical system at any fixed moment of time uniquely
determines all of its (future and past) motion.' In other words, in the phase space the position variable and the velocity variable are 
determined only by the time $t$. Therefore, we can assume that we have already a solution of Eq.(\ref{eq:inth-2})
\begin{eqnarray}
 q_i&=&q_i(t) \nonumber \\
 \dot{q_i}&=&\dot{q_i}(t),
\label{eq:curve}
\end{eqnarray}
where the solution satisfies the initial condition. We can divide the whole time domain into a group of sufficiently small domains and in these domains $q_i$ is monotone, and hence 
we can assume an inverse function $t=t(q_i)$. If $t=t(q_i)$ is substituted into the nonconservative force $\left.F_i\right|_{\gamma}$, we can assume that:
\begin{equation}
\left.F_i(q_1(t(q_i)),\cdots,q_n(t(q_i)),\dot{q}_1(t(q_i)),\cdots,\dot{q}_n(t(q_i)))\right|_{\gamma}= \mathcal{F}_i(q_i),
\label{eq:asumption}
\end{equation}
where $\mathcal{F}_i$ is a function of $q_i$ alone. In Eq.(\ref{eq:asumption}) the function $F_i$ is restricted on the curve $\gamma$, such that a new function
 $\mathcal{F}_i(q_i)$ yields. Thus we have
\begin{eqnarray}
 \int_{\gamma}F_i \dif q_i&=&\int_{q_{i0}}^{q_i}\mathcal{F}_i(q_i) \dif q_i
=W_i(q_i)-W_i(q_{i0}).
\label{eq:inth-9}
\end{eqnarray}
According to Eq.(\ref{eq:inth-9}) the function $\mathcal{F}_i$ is path independent, and therefore $\mathcal{F}_i$ can be regarded as a conservative force. 
For that Eq.(\ref{eq:asumption}) represents an identity map from the nonconservative force $F$ on the curve $\gamma$ 
to the conservative force $\mathcal{F}_i$ which is distinct from $F_i$. Eq.(\ref{eq:asumption}) is tenable only on the phase curve $\gamma$.
 Consequently the function form of $\mathcal{F}_i$ depends on the aforementioned initial condition; from other initial conditions $\mathcal{F}_i$
with different function forms will yield.

According to the physical meaning of Hamiltonian, $const$ is added to Eq.(\ref{eq:inth-8}) such
that the integral constant vanishes in Hamiltonian quantity. Hence $const=-W_i(q_{i0})$.
Substituting Eq.(\ref{eq:inth-9}) and $const=-W_i(q_{i0})$ into Eq.(\ref{eq:inth-8}), we have
\begin{equation}
\left.\hat{H}\right|_{\gamma}=H-W_i(q_i)
\label{eq:inth-10}
\end{equation}where $-W_i(q_i)$ denotes the potential of the conservative force $\mathcal{F}_i$ and $W_i(q_i)$ is equal to the sum of the work done by the 
nonconservative force $F$ and $const$. In Eq.(\ref{eq:inth-10}) $\hat{H}$ and $H$ are both functions of $q_i$ and $W_i(q_i)$ 
a function of $q_i$. 
Eq.(\ref{eq:inth-10}) and Eq.(\ref{eq:inth-8}) can be thought of as a map from the total energy of the dissipative system(\ref{eq:inth-2}) to the Hamiltonian
 of the conservative system(\ref{eq:inth-3}). Indeed, $\left.\hat{H}\right|_{\gamma}$ and the total energy differ in the constant $const=-W_i(q_{i0})$. When the
conservative system takes a different initial condition, if one does not change the function form of $\left.\hat{H}\right|_{\gamma}$, one can 
consider  $\left.\hat{H}\right|_{\gamma}$ as a Hamiltonian quantity $\hat{H}$, 
\begin{equation}
 \hat{H}=\left.\hat{H}\right|_{\gamma}=H-W_i(q_i)
\label{eq:inth-10.1}
\end{equation}

and the conservative system(\ref{eq:inth-3}) can be thought of as an entirely 
new conservative system.

\textcolor{red}{
Based on the above, the following proposition is made:
\begin{proposition}
For any nonconservative classical mechanical system and any initial condition, there exists a conservative one; the two systems share
one and only one common phase curve;  the value of the Hamiltonian of the conservative system is equal to the sum of the total energy of the nonconservative system on the aforementioned phase curve
and a constant depending on the initial condition.
\label{pro:1}
\end{proposition}}
\begin{proof} 
\textcolor{red}{
First we must prove the first part of the Proposition \ref{pro:1}, i.e. that a conservative system with Hamiltonian presented by Eq.(\ref{eq:inth-10.1})
shares a common phase curve with the nonconservative system represented by Eq.(\ref{eq:inth-2}). In other words 
the Hamiltonian quantity presented by Eq.(\ref{eq:inth-10.1}) satisfies Eq.(\ref{eq:inth-4}) under the same initial condition. 
Substituting Eq.(\ref{eq:inth-10.1}) into the left side of Eq.(\ref{eq:inth-4}), we have
\begin{eqnarray}
\frac{\partial{\hat{H}(q_i,p_i)}}{\partial {q_i}}&=&\frac{\partial H(q_i,p_i)}{\partial {q_i}}
-\frac{\partial{W_j(q_j)}}{\partial {q_i}} \nonumber\\
\frac{\partial{\hat{H}(q_i,p_i)}}{\partial {p_i}}&=&\frac{\partial H(q_i,p_i)}{\partial {p_i}}
-\frac{\partial{W_j(q_j)}}{\partial {p_i}}.
\label{eq:inth-11}
\end{eqnarray}
It must be noted that although $q_i$ and $p_i$ are considered as distinct variables in Hamilton's mechanics, we can consider $q_i$ and
$\dot{q_i}$ as dependent variables in the process of constructing of $\hat{H}$.
At the trajectory $\gamma$ we have
\begin{eqnarray}
 \frac{\partial{{W_j(q_j) }}}{\partial {q_i}}&=&
\frac{\partial{(\int_{q_{j0}}^{q_j}\mathcal{F}_j(q_j) \dif q_j+W_i(q_{i0}))}}{\partial {q_i}}
=\mathcal{F}_i(q_i) \nonumber \\
\frac{\partial{{W_j(q_j) }}}{\partial {p_i}}&=0,
\label{eq:inth-12}
\end{eqnarray}
where $\mathcal{F}_i(q_i)$ is equal to the damping force $F_i$ on the phase curve $\gamma$. Hence under the initial condition $q_0, p_0$, 
Eq.(\ref{eq:inth-4}) is satisfied. As a result, we can state that the phase curve  of Eq.(\ref{eq:inth-3}) coincides with that of 
Eq.(\ref{eq:inth-2}) under the initial condition;  and $\hat{H}$ represented by Eq.(\ref{eq:inth-10.1}) is 
the Hamiltonian of the conservative system represented by Eq.(\ref{eq:inth-3}). }

Then we must prove the second part of Proposition \ref{pro:1}: the uniqueness of the common phase curve.

 We assume that eq.(\ref{eq:inth-3}) shares two common phase curves, $\gamma_1$ and $\gamma_2$, with eq.(\ref{eq:inth-2}). 
Let a point of $\gamma_1$ at the time $t$ be $z_1$, a point of $\gamma_2$ at the time $t$  $z_2$, and 
$g^t$ the Hamiltonian phase flow of eq.(\ref{eq:inth-3}). Suppose a domain $\Omega$ at $t$ which contains only points $z_1$ and $z_2$, and $\Omega$ is 
not only a subset of the phase space of the nonconservative system(\ref{eq:inth-2}) but also that of the phase space of the conservative system(\ref{eq:inth-3}). 
Hence there exists a phase flow $\hat{g}^t$ composed of $\gamma_1$ and $\gamma_2$, and $\hat{g}^t$ is the phase flow of eq.(\ref{eq:inth-2}) restricted by $\Omega$.
According to the following Louisville's theorem\cite{Arnold1978}:
\begin{theorem}
The phase flow of Hamilton's equations preserves volume: for any region $D$ we have
\[
 volume\ of\ g^tD=volume\ of\ D
\]
where $g^t$ is the one-parameter group of transformations of phase space
\[
 g^t:(p(0),q(0))\longmapsto:(p(t),q(t))
\]
\label{Liouville}
\end{theorem}
$g^t$ preserves the volume of $\Omega$. This implies that the phase flow of eq.(\ref{eq:inth-2}) $\hat{g}^t$ preserves the volume of $\Omega$ too. 
 But the system (\ref{eq:inth-2}) is not conservative, which conflicts
with Louisville's theorem; hence only a phase curve of eq.(\ref{eq:inth-3}) coincides with that of eq.(\ref{eq:inth-2}).

\smartqed \qed
\end{proof}

\subsection{An Example in Vibration Mechanics\label{example}}

Take an $n$-dimensional oscillator with damping as an example, the governing equation of which is as below:
\begin{equation}
\ddot{\bm{q}}+\tensor{C}\dot{\bm{q}}+\tensor{K}\bm{q}=0,
\label{eq:ex2-1}
\end{equation}
where $\bm{q}=\left[q_1,\dots,q_n \right] ^T$, superscript $T$ denotes a matrix transpose,
\[
 \tensor{C}=
\left[
\begin{array}{ccc}
C_{11}&\dots&C_{1n}\\
\vdots&\ddots&\vdots\\
C_{n1}&\dots&C_{nn}
\end{array}
\right],
\tensor{K}=
\left[
\begin{array}{ccc}
K_{11}&\dots&K_{12}\\
\vdots&\ddots&\vdots\\
K_{21}&\dots&K_{22}
\end{array}
\right]
\], 
and $C_{ij}$ and $K_{ij}$ are constants.

It is complicated to solve Eq.(\ref{eq:ex2-1}). If Eq.(\ref{eq:ex2-1}) is higher dimensional, it is almost impossible to solve Eq.(\ref{eq:ex2-1}) analytically. 
Therefore we assume that a solution exists already.
\begin{equation}
 \bm{q}=\bm{q}(t)=\left[q_1(t),\dots,q_n(t)\right].
\label{eq:ex2-2}
\end{equation}
Suppose a group of inverse functions
\begin{equation}
 t=t(q_1),\dots , t=t(q_n).
\label{eq:ex2-3}
\end{equation}
As in Eq.(\ref{eq:asumption}) we can consider that the damping forces are equal to some conservative force under an initial condition
\begin{equation}
\begin{array}{ccc}
c_{11}\dot{q}_1=\varrho_{11}(q_1)&\dots&c_{1n}\dot{q}_n=\varrho_{1n}(q_1)\\
\vdots&\ddots&\vdots\\
c_{n1}\dot{q}_1=\varrho_{21}(q_n)&\dots&c_{nn}\dot{q}_n=\varrho_{nn}(q_n).
\end{array}
\label{eq:ex2-4}
\end{equation}
For convenience, these conservative forces can be thought of as elastic restoring forces: 
\begin{equation}
\begin{array}{ccc}
\varrho_{11}(q_1)=\kappa_{11}(q_1)q_1&\dots&\varrho_{1n}(q_1)=\kappa_{1n}(q_1)q_1\\
\vdots&\ddots&\vdots\\
\varrho_{n1}(q_1)=\kappa_{n1}(q_n)q_n&\dots &\varrho_{nn}(q_n)=\kappa_{nn}(q_n)q_n .
\end{array}
\label{eq:ex2-5}
\end{equation}
An equivalent stiffness matrix $\tensor{\tilde{K}}$ is obtained, which is a diagonal matrix
\begin{equation}
\tensor{\tilde{K}}_{ii}=\sum_{l=1}^n \kappa_{il}(q_l).
\label{eq:ex2-5a}
\end{equation}
Consequently an $n$-dimensional conservative system is obtained
\begin{equation}
 \bm{\ddot{q}}+(\tensor{K}+\tensor{\tilde{K}})\bm{q}=0
\label{eq:ex2-6}
\end{equation}
which shares a common phase curve with the $n$-dimensional damping system(\ref{eq:ex2-1}).
The Hamiltonian of Eqs.(\ref{eq:ex2-6}) is
\begin{equation}
 \hat{H}=\frac{1}{2}\dot{\bm{q}}^T\dot{\bm{q}}+\frac{1}{2}\bm{q}^T \tensor{K}\bm{q}+
\int_{\bm{0}}^{\bm{q}} (\tilde{\tensor{K}}\bm{q})^T \dif \bm{q},
\label{eq:ex2-7}
\end{equation}
where $\bm{0}$ is a zero vector. $\hat{H}$ in Eq.(\ref{eq:ex2-7}) is the mechanical energy of the conservative system(\ref{eq:ex2-6}), because $\int_{\bm{0}}^{\bm{q}} (\tilde{\tensor{K}}\bm{q})^T \dif \bm{q}$
 is a potential function such that $\hat{H}$ doest not 
depend on any path.

\subsection{Discussion}

Based on the above, we can outline the relationship between a dissipative mechanical system and a group of conservative systems by means of Fig. \ref{fig1}. 
The relationship can be stated from two perspectives:
\begin{figure}
\begin{center}
\includegraphics[totalheight=2.5in]{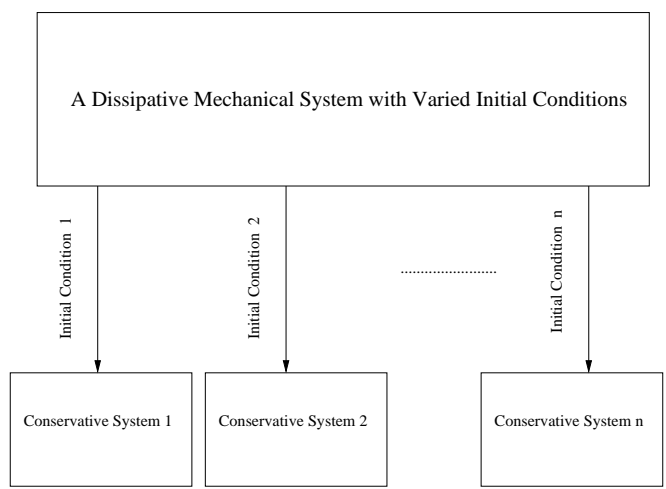}
\caption{A Dissipative Mechanical System and Conservative Systems}
\label{fig1}
\end{center}
\end{figure}

If one explains the relationship from a geometrical perspective, one can obtain Proposition \ref{pro:1}. In this paper 
the conservative systems (\ref{eq:inth-3}) and (\ref{eq:ex2-6}) are called the substituting systems. Although a 
substituting system shares a common phase curve with the original system, under other initial conditions the 
substituting system exhibits different phase curves. Therefore the phase flow of the substituting system differs 
from that of the original system, it follows that the substituting systems is not equal to 
the original system. According to Louisville's theorem (\ref{Liouville}), the phase flow of the original 
dissipative system Eq.(\ref{eq:inth-2}) certainly does not preserve its phase volume, but the phase flow of the 
substituting conservative Eq.(\ref{eq:inth-3}) does.

\textcolor{red}{
One also could explain the relationship from a mechanical perspective. 
It is known that there are non-conservative forces in a nonconservative system. The total energy of the nonconservative system consists of 
the work done by nonconservative forces. Hence the function form of the total energy depends on a phase curve i.e. under an initial condition. 
If one constrains the total energy function to a phase curve $\gamma$, the total energy function can be converted into a function of $q,p$. 
One take $\hat{H}$ consisting of this new function and a constant as a Hamiltonian quantity, such that a Hamilton's system (i.e., a 
conservative system) is obtained. Under the initial condition mentioned above, the solution curve of the 
conservative system is the same as that of the original nonconservative system; under other initial conditions 
the solution curve of the conservative is different from that of the original nonconservative system. 
Since one defines the forces(\ref{eq:asumption},\ref{eq:ex2-4},\ref{eq:ex2-5},\ref{eq:ex2-5a}) in the new system,
the Hamiltonian quantity of the conservative can be thought of as the mechanical energy of the new conservative system as Eq.(\ref{eq:ex2-7}).}

The Hamiltonians of the new conservative systems in general are not analytically integrable, unless
 the original mechanical system is integrable. The reason is that the work done by damping force
depends on the phase curve. If the system is integrable, then the
phase curve can be explicitly written out, the system has an analytical
solution, and therefore the work done by damping force can be
explicitly integrated. Subsequently, the Hamiltonian $\hat{H}$ can be explicitly expressed.
Most systems do not have an analytical solution. Despite this, the Hamilton quantity, coordinates and momentum must satisfy
 Eq.(\ref{eq:inth-3}) under a certain initial condition.  Why had Klein\cite{Klein1928} written, ''Physicists can make use of these theories only very little,
an engineers nothing at all''?  The answer: when one is seeking an analytical solution to a classical mechanics problem by 
utilizing Hamiltonian formalism, in fact one must inevitably convert the problem back to Newtonian formalism. This means that an explicit form of  Hamiltonian quantity is not necessary for classical mechanics.
What is important is the relationship between $q,p$ and the Hamiltonian quantity embodied in the Hamilton's Equation. 

According to the conclusion above, we can consider the Euler time-centered difference scheme for dissipative systems from a Hamiltonian perspective.

\section{Discussion on the Euler Time-Centered Difference Scheme}
\subsection{Introducing the Euler Time-Centered Scheme for Conservative Systems}
Feng\cite{Feng1985,Feng1989,Feng1990,Feng1991} constructed a series of symplectic difference schemes via two approaches:
 the first approach discretizes Hamilton's equation and utilizes the property of the Cayley transform to demonstrate that the map $g:\bm{z}^n\rightarrow \bm{z}^{n+1}$ is a symplectic map; 
the second approach is a so-called generating function method. Feng had represented the first approach as below:

Suppose $H$ is a differentiable function of $2n$ variables $p_1,\cdots,p_n,q_1,\cdots,q_n$, the Hamilton's equations 
are represented as:
\begin{equation}
\dot{\bm{p}}=-H_q ,\ \ \ \ \ \ \dot{\bm{q}}=H_p,
\label{eq:eq1-1}
\end{equation}
where $\bm{p}=[p_1,\cdots,p_n],\bm{q}=[q_1,\cdots,q_n],\ \ H_q=\partial H/ \partial \bm{q}$ and $H_p=\partial H/ \partial \bm{p}$. 
Let $\bm{z}=\left[\bm{p},\bm{q}\right]^T$, and Eq. (\ref{eq:eq1-1}) can be further represented as:
\begin{equation}
\dot{\bm{z}}=\tensor{J}^{-1}H_z,
\label{eq:eq1-2}
\end{equation}
where $H_z=\left[H_q,H_p \right]^T$, 
\begin{equation}
\tensor{J}=\left[ \begin{array}{cc}
 O&\tensor{I}_n\\
-\tensor{I}_n&O\\
\end{array}\right],
\label{eq:eq1-3}
\end{equation}
where $\tensor{I}_n$ and $\tensor{I}$ denote unity matrices.
If $H$ is a quadratic form:
\begin{equation}
 H(\bm{z})=\frac{1}{2}\bm{z}^TC\bm{z},\ \ \ \ C^T=C,
\label{eq:Hqudform}
\end{equation}
then canonical equations(\ref{eq:eq1-1},\ref{eq:eq1-2}) can be rewritten as:
\begin{equation}
\frac{\dif{\bm{z}}}{\dif{t}}=\tensor{B}\bm{z},\ \ \ \tensor{B}=\tensor{J}^{-1}\tensor{C}.
\label{eq:eq1-4}
\end{equation}
Paper\cite{Feng1990} gives the following definition:
\begin{definition}
A matrix $\tensor{B}$ of order $2n$ is called infinitesimal symplectic if
\[
 \tensor{JB}+\tensor{B^{T}J}=\tensor{O},
\]
where $\tensor{O}$ is a null matrix. 
All infinitesimal symplectic matrices form a Lie algebra $sp(2n)$ with commutation operation $[\tensor{A},\tensor{B}]=\tensor{AB}-\tensor{BA}$,
and $sp(2n)$ is the Lie algebra of Lie group $Sp(2n)$ known as the symplectic group.
\label{def1-1}
\end{definition}

According to this definition, $\tensor{B}=\tensor{J}^{-1}\tensor{C}$ in Eq.(\ref{eq:eq1-4}) can be considered as an infinitesimal symplectic matrix. In the paper\cite{Feng1985}
a number of symplectic schemes for Eq.(\ref{eq:eq1-4}) were proposed, one of which is the Euler time-centered scheme:
\begin{equation}
\frac{\bm{z}^{n+1}-\bm{z}^{n}}{\tau}=\tensor{B} \frac{\bm{z}^{n+1}+\bm{z}^{n}}{2}.
\label{eq:eq1-5}
\end{equation}
The transition $\bm{z}^n\rightarrow \bm{z}^{n+1}$ is given by the following linear transformation $\tensor{F}_\tau$ which coincides with its own Jacobian
\begin{equation}
\tensor{F}_\tau=\phi(-\frac{\tau}{2}\tensor{B})=(\tensor{I}-\frac{\tau}{2}\tensor{B})^{-1}(\tensor{I}+\frac{\tau}{2}\tensor{B}).
\label{eq:eq1-6}
\end{equation}
Paper\cite{Feng1990} gives the following theorem:
\begin{theorem}
If $\tensor{B}\in sp(2n)$ and $\vert \tensor{I}+\tensor{B} \vert \neq0$, then $\tensor{F}=(\tensor{I}+\tensor{B})^{-1}(\tensor{I}-\tensor{B})\in Sp(2n)$, 
the Cayley transform of $\tensor{B}$.
\label{the1-1}
\end{theorem}

According to Theorem \ref{the1-1}, $\tensor{F}_{\tau}$ can be considered as the Cayley transform of the infinitesimal symplectic matrix $-\tau\tensor{B}/2$,
 and consequently $\tensor{F}_{\tau}$ is a symplectic matrix, and $\bm{z}^n \rightarrow \bm{z}^{n+1}$ is symplectic. 

\subsection{Apply the Time-centered Difference Scheme Indirectly to Damping Oscillators}
\begin{figure}
\centering
\includegraphics[totalheight=3.5in]{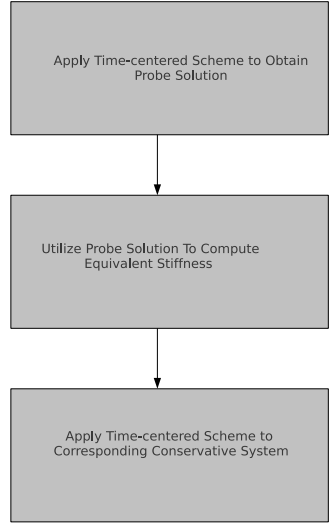}
\caption{Flowchart of applying the time-centered difference scheme indirectly to damping Oscillators in a time-step}
\label{Flow}
\end{figure}\textcolor{red}{
Based on the discussion in Section \ref{example}, we have created a flow chart (Fig.\ref{Flow}.) that shows the process of applying 
the time-centered difference scheme indirectly to the damping oscillator(\ref{eq:ex2-1}). We have defined a conservative 
system through an equivalent stiffness matrix represented  by Eq.(\ref{eq:ex2-4})(\ref{eq:ex2-5})(\ref{eq:ex2-5a}).
If one needs to formulate the equivalent stiffness matrix in numerical schemes, one must have a numerical solution at time $t^{k+1}$, 
such that an equivalent elastic restoring force can be thought of as a function of $(q_i^k+q_i^{k+1})/2$. }
\textcolor{red}{ Hence we first apply the time-centered  difference 
scheme directly to the original dissipative system. Consequently the algorithm can be depicted by Fig. \ref{Flow}.}

\textcolor{red}{The first step is to apply the time-centered difference scheme directly to Eq.(\ref{eq:ex2-1}) :}
\begin{eqnarray}
\frac{\bm{q}^{k+1}-\bm{q}^k}{\tau}&=&\frac{\bm{p}^{k+1}+\bm{p}^k}{2} \nonumber\\ 
\frac{\bm{p}^{k+1}-\bm{p}^k}{\tau}&=&-\tensor{K}\left(\frac{\bm{q}^{k+1}+\bm{q}^k}{2}\right)-\tensor{C}\frac{\dif[(\bm{q}^k+\bm{q}^{k+1})/2]}{\dif t}
\label{eq:eq2-28}
\end{eqnarray}
If we let $\dif[(\bm{q}^k+\bm{q}^{k+1})/2]/\dif t=(\bm{q}^{k+1}-\bm{q}^k)/\tau$ according to the definition of the time-centered difference scheme,  
Eq.(\ref{eq:eq2-28}) can be represented as:
\begin{eqnarray}
\left[
\begin{array}{c}
\bm{q}^{k+1}\\
\bm{p}^{k+1}\\
\end{array}
\right]=
\tensor{F}_{\tau}^1
\left[
\begin{array}{c}
\bm{q}^{k}\\
\bm{p}^{k}\\
\end{array}
\right],
\label{classical_scheme}
\end{eqnarray}
where the transition matrix is
\textcolor{red}{ 
\begin{eqnarray}
\tensor{F}_{\tau}^1=
\left[
 \begin{array}{>{\displaystyle}c>{\displaystyle}c}
 \tensor{I}&\frac{\tau}{2}\cdot \tensor{I}\\
\tau(\frac{1}{2}\cdot\tensor{K}+\frac{1}{\tau}\cdot\tensor{C}) &\tensor{I}
\end{array}
\right]
^{-1}
\left[
\begin{array}{>{\displaystyle}c>{\displaystyle}c}
 \tensor{I}&\frac{\tau}{2}\cdot \tensor{I}\\
-\tau(\frac{1}{2}\cdot\tensor{K}-\frac{1}{\tau}\cdot\tensor{C})&\tensor{I}\\
\end{array}
\right]
\label{tm_c}
\end{eqnarray}}
Obviously the difference scheme is not a symplectic scheme, because the transition matrix $F_{\tau}^1$
is not a symplectic matrix. The prerequisite proposition\cite{Feng1990} for the  proof to this point is as the following:
\textcolor{red}{
\begin{proposition}
 Matrix $\tensor{S}=\tensor{M}^{-1}\tensor{N}\in Sp(2n)$, iff $\tensor{MJM^T}=\tensor{NJN^T}$
\label{p1}
\end{proposition}}
The proof of this point is given as follows:
\begin{proof}
Since
\textcolor{red}{
 \begin{eqnarray*}
&\left[
\begin{array}{>{\displaystyle}c>{\displaystyle}c}
 \tensor{I}&\frac{\tau}{2}\cdot \tensor{I}\\
\tau(\frac{1}{2}\cdot\tensor{K}+\frac{1}{\tau}\cdot\tensor{C}) &\tensor{I}
\end{array}
\right]
\tensor{J}
\left[
\begin{array}{>{\displaystyle}c>{\displaystyle}c}
 \tensor{I}&\tau(\frac{1}{2}\cdot\tensor{K}+\frac{1}{\tau}\cdot\tensor{C})\\
\frac{\tau}{2}\cdot \tensor{I} &\tensor{I}
\end{array}
\right]
\\
=&
\left[
\begin{array}{>{\displaystyle}c>{\displaystyle}c}
 \tensor{O}&\frac{\tau^2}{2}(\frac{1}{2}\cdot\tensor{K}+\frac{1}{\tau}\cdot\tensor{C})+\tensor{I}\\
-\frac{\tau^2}{2}(\frac{1}{2}\cdot\tensor{K}+\frac{1}{\tau}\cdot\tensor{C})-\tensor{I}&\tensor{O}
\end{array}
\right]
\end{eqnarray*}}
\textcolor{red}{
 \begin{eqnarray*}
&\left[
\begin{array}{>{\displaystyle}c>{\displaystyle}c}
 \tensor{I}&\frac{\tau}{2}\cdot \tensor{I}\\
-\tau(\frac{1}{2}\cdot\tensor{K}-\frac{1}{\tau}\cdot\tensor{C}) &\tensor{I}
\end{array}
\right]
\tensor{J}
\left[
\begin{array}{>{\displaystyle}c>{\displaystyle}c}
 \tensor{I}&-\tau(\frac{1}{2}\cdot\tensor{K}-\frac{1}{\tau}\cdot\tensor{C})\\
\frac{\tau}{2}\cdot \tensor{I} &\tensor{I}
\end{array}
\right]
\\
=&
\left[
\begin{array}{>{\displaystyle}c>{\displaystyle}c}
 \tensor{O}&\frac{\tau^2}{2}(\frac{1}{2}\cdot\tensor{K}-\frac{1}{\tau}\cdot\tensor{C})+\tensor{I}\\
-\frac{\tau^2}{2}(\frac{1}{2}\cdot\tensor{K}-\frac{1}{\tau}\cdot\tensor{C})-\tensor{I}&\tensor{O}
\end{array}
\right]
\end{eqnarray*}
and according to the Proposition \ref{p1}
\begin{eqnarray*}
 F_{\tau}^1&=&
\left[
\begin{array}{>{\displaystyle}c>{\displaystyle}c}
 \tensor{I}&\frac{\tau}{2}\cdot \tensor{I}\\
\tau(\frac{1}{2}\cdot\tensor{K}+\frac{1}{\tau}\cdot\tensor{C}) &\tensor{I}
\end{array}
\right]^{-1}
\left[
\begin{array}{>{\displaystyle}c>{\displaystyle}c}
 \tensor{I}&\frac{\tau}{2}\cdot \tensor{I}\\
-\tau(\frac{1}{2}\cdot\tensor{K}-\frac{1}{\tau}\cdot\tensor{C})&\tensor{I}\\
\end{array}
\right]\\
&&\notin Sp(2n).
\end{eqnarray*}}
$F_{\tau}^1$ is not a symplectic matrix.
\smartqed
\qed 
\end{proof}

Utilizing the scheme described by Eq.(\ref{classical_scheme}) we can obtain the numerical solution $\left[\bm{q}^{k+1}, p^{k+1}\right]$ 
which is taken as a probe solution in lieu of the analytical solution, 
and then execute the second step in Fig. \ref{Flow}. Let
\textcolor{red}{
\begin{eqnarray}
\tensor{C}\cdot\dif[(\bm{q}^k+\bm{q}^{k+1})/2]/\dif t=\tensor{\tilde{K}}([(\bm{q}^k+\bm{q}^{k+1})/2),\nonumber\\
\tensor{\tilde{K}}=\left[
\begin{array}{ccc}
 \tilde{K}_{11}&\dots&0\\
\vdots&\ddots&\vdots\\
0&\dots&\tilde{K}_{nn}
\end{array}
\right]\label{approximate_solution}\\
\tilde{K}_{ii}=\sum_{j=1}^n 2C_{ij}(q^{k+1}_j-q^k_j)/(\tau(q^{k+1}_i+q^k_i)),\nonumber
\end{eqnarray}
where $\tensor{\tilde{K}}$ is the numerical approximation of the equivalent stiffness matrix(\ref{eq:ex2-5a}). Thus we would obtain 
the numerical approximation of the conservative system(\ref{eq:ex2-6}).}

Then we consider to apply the time-centered scheme to the conservative system.(\ref{eq:ex2-6}). Suppose the solution vector at the time $t^{k+1}$ is  
$\tilde{\bm{z}}^{k+1}=\left[ \tilde{\bm{q}}^{k+1}, \tilde{\bm{p}}^{k+1}\right]$
According to the time-centered difference scheme defined by Eq.(\ref{eq:Hqudform})(\ref{eq:eq1-4})(\ref{eq:eq1-5}), we 
have a time-centered scheme for the conservative system(\ref{eq:ex2-6}):
\begin{eqnarray}
 \frac{\tilde{\bm{q}}^{k+1}-\bm{q}^k}{\tau}&=&\frac{\tilde{\bm{p}}^{k+1}+\bm{p}^k}{2} \nonumber\\ 
\frac{\tilde{\bm{p}}^{k+1}-\bm{p}^k}{\tau}&=&-(\tensor{K}+\tensor{\tilde{K}})\left(\frac{\tilde{\bm{q}}^{k+1}+\bm{q}^k}{2}\right).
\label{eq:indirect_scheme}
\end{eqnarray}
The time-centered scheme above can be represented as
\begin{equation}
 \left[
\begin{array}{c}
\tilde{\bm{q}}^{k+1}\\
\tilde{\bm{p}}^{k+1}\\
\end{array}
\right]=
\tensor{F}_{\tau}^2
\left[
\begin{array}{c}
{\bm{q}}^{k}\\
{\bm{p}}^{k}\\
\end{array}
\right].
\label{symplectic_scheme}
\end{equation}
\textcolor{red}{
where 
\begin{eqnarray}
\tensor{F}_{\tau}^2&=&
\left[
 \begin{array}{>{\displaystyle}c>{\displaystyle}c}
 \tensor{I}&-\frac{\tau}{2} \cdot\tensor{I}\\
\frac{\tau}{2}(\tensor{K}+\tensor{\tilde{K}}) &\tensor{I}\\
\end{array}
\right]^{-1} \cdot\nonumber\\
&&\left[
\begin{array}{>{\displaystyle}c>{\displaystyle}c}
 \tensor{I}&\frac{\tau}{2}\cdot \tensor{I}\\
-\frac{\tau}{2}(\tensor{K}+\tensor{\tilde{K}})&\tensor{I}\\
\end{array}
\right]
\label{tm_s}
\end{eqnarray}
According to Eq.(\ref{symplectic_scheme}) and Eq.(\ref{tm_s}), the map $ \bm{z}^k\rightarrow \tilde{\bm{z}}^{k+1}$ must be symplectic, because $\tensor{F}_{\tau}^2$
must be a symplectic matrix. The proof of this point is as follows:
\begin{proof}
\begin{eqnarray*}
\tensor{F}_{\tau}^2&=&\left( \tensor{I}+
\left[\begin{array}{>{\displaystyle}c>{\displaystyle}c}
\tensor{O}&-\frac{\tau}{2} \cdot\tensor{I}\\
\frac{\tau}{2}(\tensor{K}+\tensor{\tilde{K}})&\tensor{O}
\end{array}\right]\right)^{-1}
\\
&&\left( \tensor{I}-
\left[\begin{array}{>{\displaystyle}c>{\displaystyle}c}
\tensor{O}&-\frac{\tau}{2} \cdot\tensor{I}\\
\frac{\tau}{2}(\tensor{K}+\tensor{\tilde{K}})&\tensor{O}
\end{array}\right]
\right)
\end{eqnarray*}
Since 
\begin{eqnarray*}
&&\tensor{J}
\left[\begin{array}{>{\displaystyle}c>{\displaystyle}c}
\tensor{O}&-\frac{\tau}{2} \cdot\tensor{I}\\
\frac{\tau}{2}(\tensor{K}+\tensor{\tilde{K}})&\tensor{O}
\end{array}\right]
+\\
&&\left[\begin{array}{>{\displaystyle}c>{\displaystyle}c}
\tensor{O}&-\frac{\tau}{2} \cdot\tensor{I}\\
\frac{\tau}{2}(\tensor{K}+\tensor{\tilde{K}})&\tensor{O}
\end{array}\right]^{T}\tensor{J}=\tensor{O}
\end{eqnarray*}
and according to Definition \ref{def1-1}
\[
 \left[\begin{array}{>{\displaystyle}c>{\displaystyle}c}
\tensor{O}&-\frac{\tau}{2} \cdot\tensor{I}\\
\frac{\tau}{2}(\tensor{K}+\tensor{\tilde{K}})&\tensor{O}
\end{array}\right]
\]
is an infinitesimal symplectic matrix. 
Therefore, according to Theorem \ref{the1-1}, $\tensor{F}^2_\tau$ is a symplectic matrix.
\end{proof}}

Finally, by substituting $\bm{z}^k=\left[ \bm{q}^k, \bm{p}^k\right]$ and Eqs.(\ref{approximate_solution}) into Eq.(\ref{symplectic_scheme}) and Eq.(\ref{tm_s}),
we can obtain the numerical solution $\tilde{\bm{z}}^{k+1}$. Then if we repeat the process that consists of Eq.(\ref{classical_scheme})(\ref{approximate_solution})(\ref{symplectic_scheme}), 
we would get a series of numerical solutions for Eq.(\ref{eq:ex2-1}) that are canonical for its substituting conservative system.

\textcolor{red}{From the derivation above, we can obtain the numerical solution $\bm{z}^{k+1}$ by direct application to the dissipative system(\ref{eq:ex2-1}) and the numerical solution $\tilde{\bm{z}}^{k+1}$ 
by indirect application.
Although $\tensor{F}^1_\tau\neq \tensor{F}^2_\tau$, from Eq.(\ref{classical_scheme}), Eq.(\ref{tm_c}), Eq.(\ref{approximate_solution}) and Eq.(\ref{eq:indirect_scheme}) 
we can derive
\begin{eqnarray}
\left[
\begin{array}{c}
\tilde{\bm{q}}^{k+1}\\
\tilde{\bm{p}}^{k+1}\\
\end{array}
\right]&=&
\left[
\begin{array}{c}
{\bm{q}}^{k+1}\\
{\bm{p}}^{k+1}\\
\end{array}
\right] \nonumber\\&=&
\left[
\begin{array}{c}
\dfrac{4\bm{q}^k-\tau^2\tensor{K}\bm{q}^k+2\tau\tensor{C}\bm{q}^k+4\tau\bm{p}^k}{4+\tau^{2}\tensor{K}+2\tau \tensor{C}}\\
-\dfrac{4\tau \tensor{K}\bm{q}^k+\tau^2\tensor{K}\bm{p}^{k}+2\tau\tensor{C}\bm{p}^k-4\bm{p}^k}{4+\tau^2\tensor{K}+2\tau\tensor{C}}\\
\end{array}
\right].
\label{same_scheme}
\end{eqnarray}
According to Eq.(\ref{same_scheme}), the result of the transition $\bm{z}^k\rightarrow \bm{z}^{k+1}$ is exactly identical to 
that of $\bm{z}^k\rightarrow \tilde{\bm{z}}^{k+1}$, and hence we have $\bm{z}^{k+1}=\tilde{\bm{z}}^{k+1}$. It is a known fact, 
that the linear transition} \textcolor{red}{ matrix between vectors $\bm{z}^k$ and $\bm{z}^{k+1}$ should not be unique; hence $F^1_\tau$ may not be equal to $F^2_\tau$.
 Therefore, we can say that the time centered difference scheme(\ref{classical_scheme}) for a dissipative mechanical system is in fact also 
a symplectic scheme for the substituting conservative system. Consequently, one does not need to execute the second and third steps in Fig. \ref{Flow}.}
\subsection{Numerical Examples}
\begin{figure}
\begin{center}
\includegraphics[totalheight=2.5in]{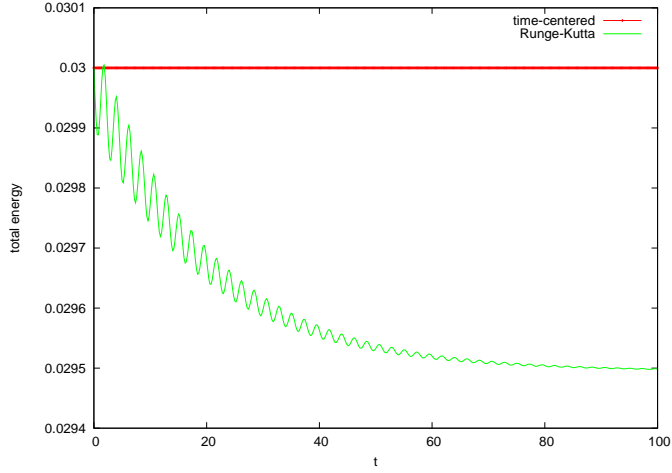}
\caption{Total energy behavior of integrators for an one-dimensional damping oscillator}
\label{result_1n}
\end{center}
\end{figure}
\begin{figure}
\begin{center}
\includegraphics[totalheight=2.5in]{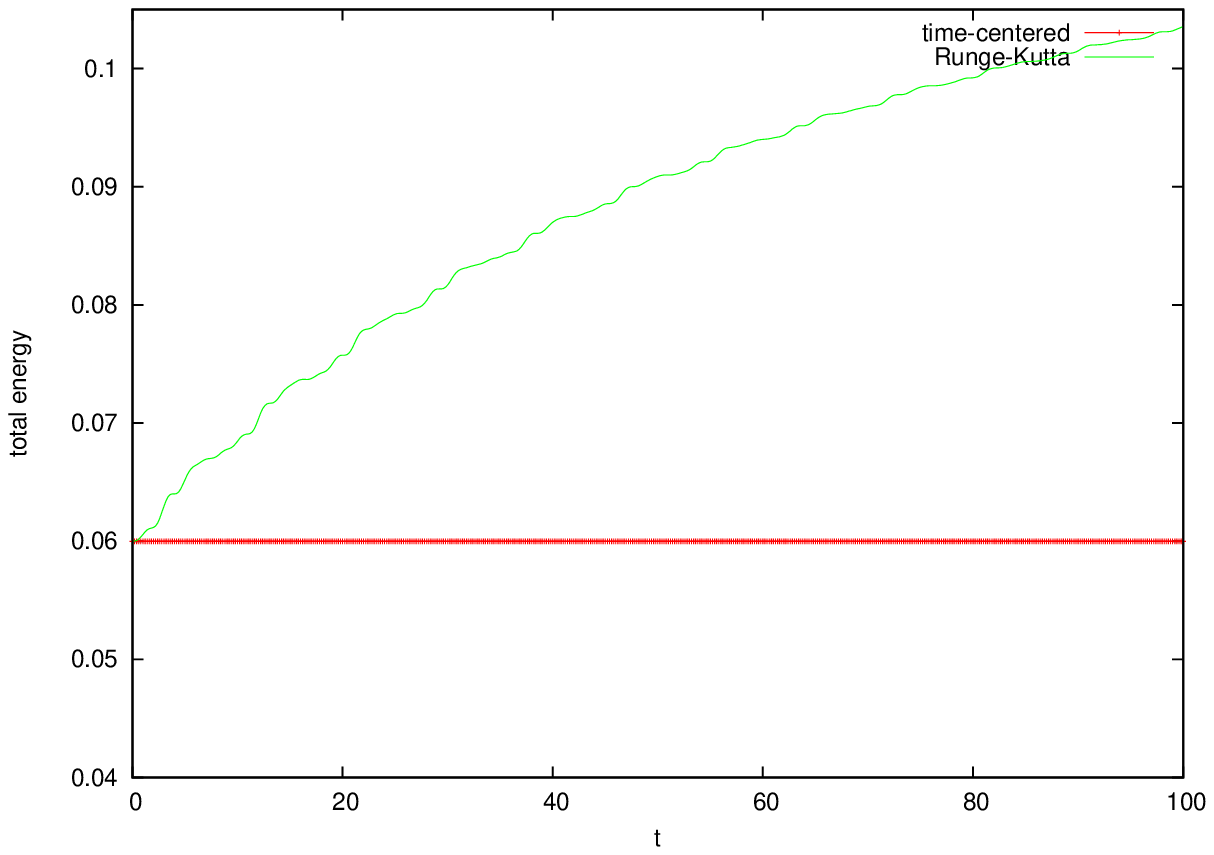}
\caption{Total energy behavior of integrators for a two-dimensional damping oscillator}
\label{result_2n}
\end{center}
\end{figure}
\begin{figure}
\begin{center}
\includegraphics[totalheight=2.5in]{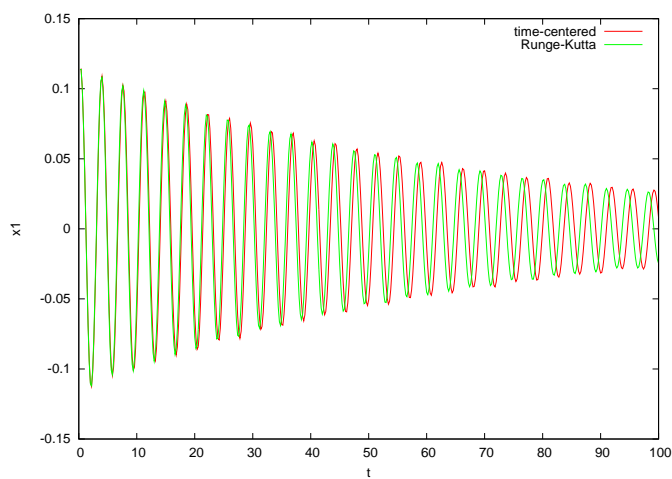}
\caption{1st displacement of a two-dimensional damping oscillator}
\label{result_x1}
\end{center}
\end{figure}

Although the time-centered difference scheme is used widely, the total energy behavior of this scheme for dissipative systems 
has not been illustrated via numerical examples. We now have two numerical examples to 
show the total energy behavior for damping systems.

The first numerical example involves an one-dimensional damping oscillator 
\[\ddot{x}+2x+0.05\dot{x}=0
\]
with initial values $x_0=0.1, \ \ \dot{x}_0=0.2$. The time-centered scheme with step-size=0.2 and the 4th-Runge-Kutta method 
with  step-size=0.2 are used to compute the one-dimensional problem. The total energy result is shown in Fig. \ref{result_1n}.

The second numerical example involves a two-dimensional damping oscillator
\begin{eqnarray*}
\ddot{x_1}+3x_1+0.03\dot{x_1}-0.01\dot{x_2}=0\\
\ddot{x_2}+3x_2-0.01\dot{x_1}+0.01\dot{x_2}=0,
\end{eqnarray*}
with initial values $x_1|_{t=0}=0.1, \ \ \dot{x}_1|_{t=0}=0.1,\ \ x_2|_{t=0}=0.2, \ \ \dot{x}_2|_{t=0}=0.2$. The same method 
as earlier has been employed to compute this problem. The numerical 
result of the 1st displacement is shown in Fig.\ref{result_x1}, and the total energy result in Fig. \ref{result_2n}.

As shown in Fig. \ref{result_1n} and Fig. \ref{result_2n}, the Euler time-centered difference scheme can preserve the total energy, 
\textcolor{red}{
the primary reason being is that the foundation of the Euler time-}\textcolor{red}{centered difference scheme is a Hamilton's system, 
the Hamiltonian quantity of which is the sum of the total energy and a constant.} By comparing Fig. \ref{result_2n} and 
Fig. \ref{result_x1}, one will find that the period of the result of the time-centered scheme is 
shorter than that of the result of the Runge-Kutta method. The reason is clear: the Runge-Kutta method would cause artificial energy growth.
This result can be considered as a generalization of the conclusion in the paper\cite{xingyufeng2007}, which asserts that the 
Euler time-centered scheme preserves the mechanical energy of conservative oscillators. By comparing the results in this 
paper and the work in the paper\cite{Kane2000}, we can identify the difference. As presented in Fig. 6.1 and Fig. 6.2 in 
the paper \cite{Kane2000}, 'The variational integrators simulate energy decay, unlike standard methods such as Runge-Kutta.', 
there is no accurate criterion of numerical integration methods for dissipative system just as the accurate criterion for conservative 
systems which is a horizontal line(see Fig. 4.1 and Fig. 4.2 in the paper\cite{Kane2000}), its rationale has have been presented 
in Sec.\ref{Introduction}.

\section{conclusions}
We can conclude that a dissipative mechanical system has such
properties: for any nonconservative classical mechanical system and any initial condition, 
there exists a conservative one, the two systems share
one and only one common phase curve;  the Hamiltonian of the conservative 
system is the sum of the total energy of the nonconservative system on the aforementioned phase curve
and a constant depending on the initial condition.  We can further conclude,
that a dissipative problem can be reformulated as an infinite number
of non-dissipative problems, one corresponding to each phase curve
of the dissipative problem. One can avoid having to change the
definition of the canonical momentum in the Hamilton formalism,
because under a certain initial condition the motion of one of the
group of conservative systems is the same as the original dissipative
system.

From a Hamiltonian perspective, this paper has revealed that the transition matrix of $\bm{z}^k\rightarrow \bm{z}^{k+1}$ 
is not unique, the transition matrix may be non-symplectic or symplectic, the numerical solution of the original dissipative obtained through 
the time-centered difference scheme system is equal to that of the substituting conservative system obtained through 
a time-centered difference scheme.
In addition, we have discovered that the time-centered scheme preserves the total energy of dissipative systems.
Based on the above, we might explain why the time-centered difference scheme is more accurate than an unsymplectic one 
for dissipative problem. For this reason we might be able to choose and construct better algorithms.

\bibliography{mybib.bib}

\end{document}